\documentclass[aps,showpacs,twocolumn,eqsecnum]{revtex4}

\usepackage{latexsym}
\usepackage{amssymb}
\usepackage{amsmath}
\usepackage{epsfig}

\begin{document}


\title{Are gauge shocks really shocks?}


\author{Miguel Alcubierre}

\affiliation{Instituto de Ciencias Nucleares, Universidad Nacional
Aut{\'o}noma de M{\'e}xico, A.P. 70-543, M{\'e}xico D.F. 04510, M{\'e}xico}


\date{March, 2005}


\begin{abstract}
  The existence of gauge pathologies associated with the Bona-Masso
  family of generalized harmonic slicing conditions is proven for the
  case of simple 1+1 relativity.  It is shown that these gauge
  pathologies are true shocks in the sense that the characteristic
  lines associated with the propagation of the gauge cross, which
  implies that the name ``gauge shock'' usually given to such
  pathologies is indeed correct.  These gauge shocks are associated
  with places where the spatial hypersurfaces that determine the
  foliation of spacetime become non-smooth.
\end{abstract}




\maketitle


\section{Introduction}
\label{sec:intro}

When studying the dynamical evolution of spacetime, it is important to
choose coordinates that allow one to cover as large a region of
spacetime as possible without becoming pathological.  In the 3+1
approach, this choice of coordinates naturally separates in two
different aspects: The choice of a time coordinate, that is, of a
foliation of spacetime into spatial hyper-surfaces (also known as the
``slicing''), associated with the lapse function $\alpha$, and the
choice of the way in which the spatial coordinates (the ``time
lines'') propagate from one hyper-surface to the next, associated with
the shift vector $\beta^i$.

With respect to the choice of slicing, many different ways to choose
the lapse function are possible. One can, for example, specify the
lapse directly as a function of the geometric variables.
Alternatively, one can obtain the lapse as a solution of some specific
differential equation.  Elliptic slicing conditions are typically
obtained when one enforces some geometric condition on the spatial
hypersurfaces.  An example of an elliptic slicing condition is
``maximal slicing''~\cite{Smarr78b}, which requires that the spatial
volume elements remain constant during the evolution and has strong
singularity avoiding properties, making it particularly well suited
for studies of black hole spacetimes.

Another possibility is to specify an evolution equation for the lapse
function and evolve this function in time as just another dynamical
quantity.  This last approach has the advantage that it is much easier
computationally to evolve the lapse than to solve an elliptic equation
for it.  One particular family of evolution type slicing conditions
was introduced by Bona and Masso as a generalization of the harmonic
time coordinate condition~\cite{Bona94b}.  This family has the
important property of allowing one to construct strongly hyperbolic
formulations of the Einstein evolution equations that include the
slicing condition.  Also, some members of the Bona-Masso family have
been shown to mimic the singularity avoiding properties of maximal
slicing.

In 1997, I studied different members of the Bona-Masso family and
found that, unless a specific condition was imposed, gauge pathologies
could easily develop~\cite{Alcubierre97a}.  I called those pathologies
``gauge shocks'' as they appeared as discontinuities in the lapse and
metric functions that developed from smooth initial data.
In~\cite{Alcubierre02b} I strengthened the case for the existence of
these gauge shocks by providing a purely kinematic argument
independent of the Einstein equations, and more recently
in~\cite{Reimann:2004wp} gauge shocks have again been predicted
together with blow-ups associated with constraint violation.  Still,
there have been some doubts in the numerical relativity community
about both the relevance and the reality of gauge shocks.  Recently
it has even been claimed by Bona~{\em et al.}~\cite{Bona:2004ma} that
genuine gauge shocks are completely discarded.

The purpose of this paper is therefore to provide a rigorous proof in
the case of simple 1+1 dimensional relativity that gauge pathologies
associated with the Bona-Masso family of slicing conditions do
develop, and that these pathologies are genuine shocks in the
following sense: They correspond to discontinuities in the solutions
of the underlying hyperbolic differential equations that develop from
smooth initial data and are associated with the crossing of the
characteristic lines.

It is important to stress the fact that the term ``shock'' will be
used here strictly in this restricted sense of the crossing of the
characteristic lines.  In hydrodynamics, the word shock is associated
also with the subsequent propagation of these discontinuities (``shock
waves'').  However, since at a discontinuity the differential
equations break down, one can only talk about so-called ``weak''
solutions. Such weak solutions are not unique, and one needs to
specify an extra physical principle known as an ``entropy condition''
to identify the correct solution.  In the case of gauge shocks,
however, once the discontinuity develops our gauge has in fact broken
down. It is unlikely that one can find an entropy condition in this
case as the gauge can be chosen arbitrarily, so any weak solution
should be acceptable in principle (though they would all be
singular). Notice also that we are really interested in finding how
such gauge shocks can best be avoided since discontinuities in the
gauge variables are clearly undesirable. We are not interested in
propagating these discontinuities.~\footnote{It has been suggested
that because of the fact that in this case there is no entropy
condition, the name ``gauge shock'' in not very appropriate and one
should just call these singularities ``gauge pathologies''.  However,
this name is simply too general as gauge pathologies can develop in
many different ways. The pathologies discussed here are of a very
specific type, and they develop precisely as shocks do, {\em i.e.}
because the characteristic lines associated with the propagation of
the gauge cross.}

This paper is organized as follows.  In Sec.~\ref{sec:1p1} I review
the basic equations of 1+1 relativity, and show that they can in fact
be written as a system of conservation laws, both in terms of the
geometric variables and in terms of the eigenfields.  My main argument
for the existence of gauge shocks is presented in
Sec.~\ref{sec:shocks}, first showing that blow ups of the eigenfields
can develop in finite time, and then showing that the characteristic
lines associated to the gauge propagation do cross.  I also show a
numerical example showing how a guage shock develops.  I conclude in
Sec~\ref{sec:conclusions}.  Appendix~\ref{sec:appendix_initial}
describes the initial data used for the numerical simulations, and
finally Appendix~\ref{sec:appendix_bona} counters the argument given
by Bona {\em et al.} in~\cite{Bona:2004ma} to claim that gauge shocks
do not occur.


\section{Einstein equations in 1+1 dimensions}
\label{sec:1p1}

Let us consider vacuum general relativity in one spatial dimension. It
is well known that in such a case the gravitational field is trivial
and there are no true dynamics.  However, one can still have
nontrivial gauge dynamics that can be used as a simple example of the
type of behavior one can expect in the higher dimensional case.

As slicing condition we will use the Bona-Masso family of generalized
harmonic slicing conditions~\cite{Bona94b}
\begin{equation}
\partial_t \alpha = - \alpha^2 f(\alpha) K \; ,
\label{eq:BonaMasso}
\end{equation}
with $K = K_x^x$ the trace of the extrinsic curvature.

The ADM~\cite{Arnowitt62,York79} evolution equations in the 1+1 case
can be written in first order form as
\begin{eqnarray}
\partial_t \alpha &=& - \alpha^2 f K \; ,
\label{eq:alphadot} \\
\partial_t g &=& - 2 \alpha g K \; ,
\label{eq:gdot}
\end{eqnarray}
and
\begin{eqnarray}
\partial_t D_\alpha  + \partial_x \left( \alpha f K \right) &=& 0 \; ,
\label{weq:Dadot} \\
\partial_t D_g  + \partial_x \left( 2 \alpha K \right) &=& 0 \; ,
\label{eq:Dgdot} \\
\partial_t K + \partial_x \left( \alpha D_\alpha /g \right)
&=& \alpha \left( K^2 - D_\alpha D_g / 2g \right) \; ,
\label{eq:Kdot}
\end{eqnarray}
where we have defined $g := g_{xx}$, $D_\alpha :=
\partial_x \ln{\alpha}$ and \mbox{$D_g := \partial_x \ln{g}$}.

Before doing an analysis of the characteristic structure of this
system of equations, it is important to notice that the evolution
equation for $K$ can in fact be rewritten as a conservation law in the
following way
\begin{equation}
\partial_t \left( g^{1/2} K \right)
+ \partial_x \left( \alpha D_\alpha / g^{1/2} \right) = 0 \; .
\label{eq:Kdot_cons}
\end{equation}
If we now define the vector $\vec{v}:=(D_\alpha, D_g, \tilde{K} )$,
with \mbox{$\tilde{K} := g^{1/2} K$}, then the evolution equations for
the first order variables can be written as a conservative system of
the form
\begin{equation}
\partial_t \vec{v} + \partial_x \left( {\bf M} \: \vec{v} \right) = 0 \; ,
\label{eq:vdot}
\end{equation}
with the characteristic matrix ${\bf M}$ given by:
\begin{equation}
{\bf M} = \left( \begin{array}{ccc}
0 & 0 & \alpha f / g^{1/2} \\
0 & 0 & 2 \alpha / g^{1/2} \\
\alpha / g^{1/2} & 0 & 0
\end{array} \right) \; .
\label{eq:matrixM}
\end{equation}
The characteristic matrix has the following eigenvalues
\begin{equation}
\lambda_0 = 0 \; , \qquad
\lambda_\pm = \pm \alpha \left( f / g \right)^{1/2} \; ,
\label{eq:eigenvalues}
\end{equation}
with corresponding eigenvectors
\begin{equation}
\vec{e}_0 = \left( 0 , 1 , 0 \right) \; , \qquad
\vec{e}_\pm = \left( f , 2 , \pm f^{1/2} \right) \; .
\label{eq:eigenvectors}
\end{equation}
Since the eigenvalues are real for $f>0$ and the eigenvectors are
linearly independent, the system~(\ref{eq:vdot}) is strongly
hyperbolic. The eigenfunctions are given by
\begin{equation}
\vec{\omega} = {\bf R}^{-1} \vec{v} \; ,
\end{equation}
with ${\bf R}$ the matrix of column eigenvectors.  We find (using an
adequate choice of normalization, see below)
\begin{equation}
\omega_0 = D_\alpha / f - D_g / 2 \; , \qquad
\omega_\pm = \tilde{K} \pm D_\alpha / f^{1/2} \; ,
\label{eq:eigenfunc}
\end{equation}
which can be easily inverted to give
\begin{eqnarray}
\tilde{K} &=& \frac{\left( \omega_+ + \omega_- \right)}{2} \; , \\
D_\alpha &=& \frac{f^{1/2} \: \left( \omega_+ - \omega_- \right)}{2} \; , \\
D_g &=& \frac{\left( \omega_+ - \omega_- \right)}{f^{1/2}} - 2 \omega_0 \; .
\end{eqnarray}

It is important to notice that with the eigenfunctions scaled as
above, their evolution equations also turn out to be conservative and
have the simple form:
\begin{equation}
\partial_t \vec{\omega}
+ \partial_x \left( {\bf \Lambda} \: \vec{\omega} \right) = 0 \: ,
\label{eq:omegadot}
\end{equation}
with ${\bf \Lambda} = {\rm diag} \left\{ \lambda_i \right\}$.  If,
however, the eigenfunctions are rescaled in the way $\omega'_i =
F_i(\alpha,g) \: \omega_i$, then the evolution equations for the
$\omega'_i$ will in general no longer be conservative and non-trivial
sources will be present.  The important point is that there is in fact
one normalization in which the equations are conservative, namely the
one given in~(\ref{eq:eigenfunc}).


\section{Gauge shocks in 1+1 relativity}
\label{sec:shocks}

There are two different ways in which one can see that the evolution
equations derived in the previous section develop singular solutions.
Let us start by looking at the evolution equations for the traveling
eigenfunctions $\omega_\pm$
\begin{equation}
\partial_t \omega_\pm + \partial_x \left( \lambda_\pm \omega_\pm
\right) = 0 \; .
\end{equation}
We now rewrite these equations as
\begin{equation}
\partial_t \omega_\pm + \lambda_\pm \partial_x \omega_\pm =
- \omega_\pm \partial_x \lambda_\pm \; .
\end{equation}
Using the expressions for $\lambda_\pm$, and denoting \mbox{$f' \equiv
df/d\alpha$}, one finds
\begin{eqnarray}
\partial_x \lambda_\pm &=& \mp \frac{\alpha f^{1/2}}{2 g^{3/2}} \: \partial_x g
\pm \frac{f^{1/2}}{g^{1/2}} \left( 1 + \frac{\alpha f'}{2 f} \right)
\partial_x \alpha \nonumber \\
&=& \lambda_\pm \left[ \left( f + \frac{\alpha f'}{2} \right)
\: \frac{D_\alpha}{f} - \frac{D_g}{2} \right] \nonumber \\
&=& \lambda_\pm \left[ \left( f - 1 + \frac{\alpha f'}{2} \right)
\frac{\omega_+ - \omega_-}{2 f^{1/2}} + \omega_0 \right] \: , \hspace{5mm}
\end{eqnarray}
and finally 
\begin{eqnarray}
\partial_t \omega_\pm + \lambda_\pm \partial_x \omega_\pm = \hspace{40mm}
\nonumber \\
\lambda_\pm \omega_\pm \left[ \left( 1 - f - \frac{\alpha f'}{2} \right)
\frac{\omega_+ - \omega_-}{2 f^{1/2}} - \omega_0 \right] \: .
\label{eq:omegadot_long}
\end{eqnarray}

Assume now that we are in a region such that \mbox{$\omega_0 =
\omega_- = 0$}.  It is clear that $\omega_0$ will not be excited,
since it does not evolve, nor will $\omega_-$ be excited, since from
the equation above we see all that its sources vanish.  The evolution
equation for $\omega_+$ then simplifies to
\begin{equation}
\partial_t \omega_+ + \lambda_+ \partial_x \omega_+
= \frac{\lambda_+}{2 f^{1/2}} \left( 1 - f
- \frac{\alpha f'}{2} \right) \omega_+^2 \; .
\end{equation}
This equation shows that $\omega_+$ will blow up along its
characteristics unless the term in parenthesis vanishes:
\begin{equation}
1 - f - \frac{\alpha f'}{2} = 0 \; .
\label{eq:shockcondition}
\end{equation}
The last condition has been derived several times
before~\cite{Alcubierre97a,Alcubierre02b,Reimann:2004wp}, and can be
easily integrated to give
\begin{equation}
f(\alpha) = 1 + k/\alpha^2 \; ,
\label{eq:alphanoshock}
\end{equation}
with $k$ an arbitrary constant.  Notice that harmonic slicing given by
$f=1$ is of this form, but $f={\rm constant}\neq1$ is not.  Notice
also that ``1+log'' slicing for which \mbox{$f=2/\alpha$}, though not
of the form~(\ref{eq:alphanoshock}), nevertheless satisfies
condition~(\ref{eq:shockcondition}) at places where $\alpha=1$.

In some cases it is even possible to predict exactly when a blow up
will occur.  In order to see this we first define the rescaled
eigenfunctions $\Omega_\pm := \alpha \omega_\pm/g^{1/2}$. For their
evolution equations we now find
\begin{eqnarray}
\partial_t \Omega_\pm + \lambda_\pm \partial_x \Omega_\pm &=&
\left( 1 - f - \frac{\alpha f'}{2} \right) \frac{\Omega^2_\pm}{2} \nonumber \\
&+&  \left( 1 - f + \frac{\alpha f'}{2} \right)
\frac{\Omega_\pm \Omega_\mp}{2} \: . \hspace{10mm}
\end{eqnarray}
Notice that with this new scaling, all contributions from $\omega_0$
to the sources have disappeared.  If we now assume that we have
initial data such that $\Omega_-=0$, then the evolution equation for
$\Omega_+$ simplifies to
\begin{equation}
\partial_t \Omega_+ + \lambda_+ \partial_x \Omega_+ = \left( 1 - f
- \frac{\alpha f'}{2} \right) \frac{\Omega^2_+}{2} \: ,
\end{equation}
which can be rewritten as
\begin{equation}
\frac{d \Omega_+}{dt} = \left( 1 - f - \frac{\alpha f'}{2} \right)
\frac{\Omega^2_+}{2} \; ,
\end{equation}
with $d/dt$ the derivative along the characteristic.  It is clear that
we have the same condition for avoiding blow-ups as before.  But the
last equation has a very important property: for constant $f$ the
coefficient of the quadratic source term is itself also constant. In
that case the equation can be easily integrated to find (assuming $f
\neq 1$)
\begin{equation}
\Omega_+ = \frac{\Omega_+^0}{1 - (1 - f) \: \Omega_+^0 \: t / 2} \: ,
\end{equation}
where $\Omega_+^0 \equiv \Omega_+(t=0)$.  The solution will then blow
up at a finite time given by $t^* = 2/[(1-f) \: \Omega_+^0]$.
Clearly, this time will be in the future if $(1-f) \: \Omega_+^0 >0$,
otherwise it will be in the past.  Since in general $\Omega_+^0$ will
not be constant in space, the first blow-up will occur at the time
\begin{equation}
\mbox{$T^* = 2/[(1-f) \: \max(\Omega_+^0(x))]$} \; .
\label{eq:blowuptime}
\end{equation}

\begin{figure}
\epsfxsize=100mm
\epsfysize=70mm
\epsfbox{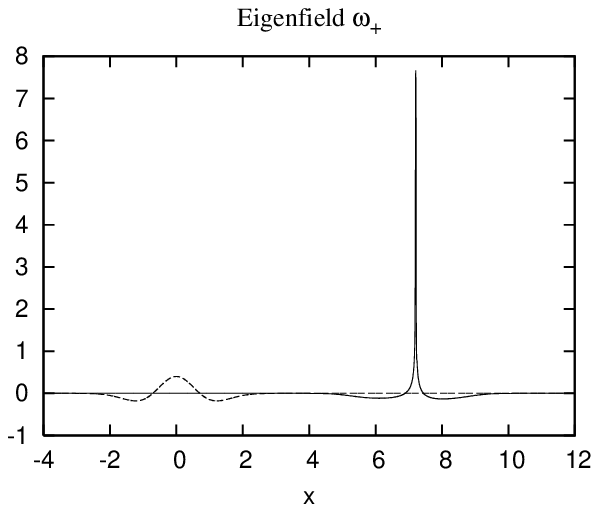}
\caption{Evolution of the eigenfunction $\omega_+$ for initial data
representing a pulse traveling to the right.  The dotted line shows
the initial data, and the solid line the numerical solution at $t=10$.
Notice how $\omega_+$ has developed a large spike.}
\label{fig:omega}
\end{figure}

\begin{figure}
\epsfxsize=100mm
\epsfysize=70mm
\epsfbox{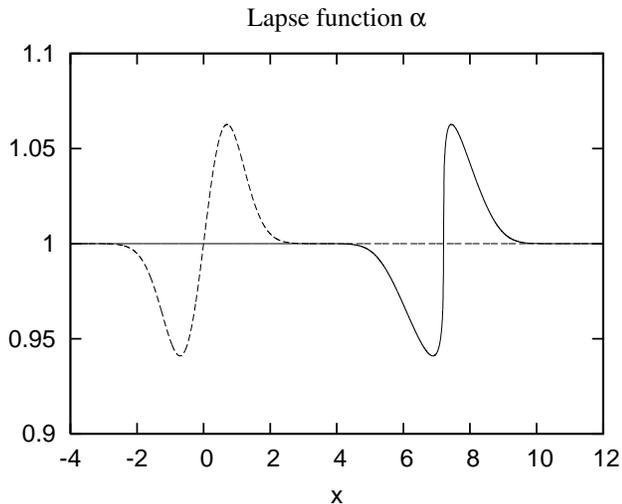}
\caption{Evolution of the lapse function $\alpha$ for initial data
representing a pulse traveling to the right.  The dotted line shows
the initial data, and the solid line the numerical solution at $t=10$.
Notice how $\alpha$ has developed a sharp gradient.}
\label{fig:lapse}
\end{figure}

Figures~\ref{fig:omega} and~\ref{fig:lapse} show the numerical
evolution of the eigenfield $\omega_+$ and the lapse function
$\alpha$, in a case when $f=1/2$, for initial data such that
$\omega_-=0$ and using a resolution of $\Delta x = 0.003125$ (see
Appendix A for details on how to construct such initial data).  For the
initial data used here, according to~(\ref{eq:blowuptime}) a blow-up
is expected at time $T^* = 9.98$.  The plots show both the initial
data (dotted lines) and the numerical solution at time \mbox{$t=10$}
(solid lines), just after the expected blow-up.  We clearly see how
the eigenfield $\omega_+$ has developed a large spike, while the lapse
has developed a sharp gradient (the solution does not become infinite
because the numerical method used has some inherent
dissipation~\footnote{For those readers interested in the numerical
details, the code uses a method of lines integration in time, with
either 3-step iterative Crank-Nicholson or 4th order Runge-Kutta. For
the spatial differentiation the code uses a slope limiter method of
van Leer's type, see~\cite{Leveque92}.  No artificial dissipation is
introduced other than the inherent dissipation of the slope limiter
itself.}).

\begin{figure}
\epsfxsize=100mm
\epsfysize=70mm
\epsfbox{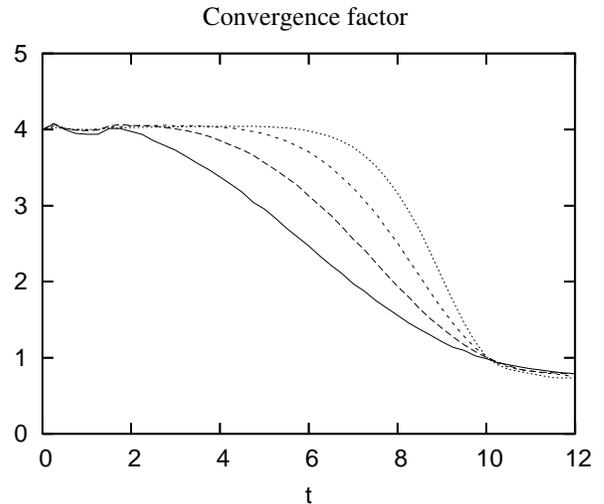}
\caption{Convergence factors for the constraint \mbox{$C_\alpha$} for
runs done at the five different resolutions: $\Delta x=$ 0.05, 0.025,
0.0125, 0.00625, 0.003125. As resolution increases, the convergence
factors approach the expected value of 4 for $t<10$ (the curves move
up), but after that time they drop sharply indicating loss of
convergence.}
\label{fig:convergence}
\end{figure}

If one repeats the simulation at different resolutions one finds that
the numerical solution converges up to a time $t \sim 10$, and after
that convergence fails, indicating that even though the numerical
solution continues past this time, we are no longer solving the
original differential equations. To see this we can consider the
convergence of the constraint \mbox{$C_\alpha := D_\alpha - \partial_x
\ln \alpha$}.  Numerically this constraint will not vanish, but it
should converge to zero as the resolution is increased.  Define now
the convergence factor as the ratio of the r.m.s norm of $C_\alpha$
for a run at a given resolution and another run at twice the
resolution.  Figure~\ref{fig:convergence} shows a plot of the
convergence factors as a function of time for runs done at five
different resolutions: $\Delta x =$ 0.05, 0.025, 0.0125, 0.00625,
0.003125.  Since the numerical code is second order, the convergence
factors should be close to 4. The figure shows that as the resolution
is increased the convergence factors approach the expected value of 4
for $t<10$ (the lines move up), but after that time they drop below 1
indicating loss of convergence.

We have then found that blow-ups do develop for the Bona-Masso family
of slicing conditions whenever~(\ref{eq:shockcondition}) does not
hold, and we can even predict the precise time of blow up formation
when $f(\alpha)$ is a constant.  The question now is whether these
blow ups are genuine shocks or not.  To answer this question let us
now consider the evolution of the eigenspeeds themselves along their
corresponding characteristic lines.  From~(\ref{eq:eigenvalues}) we
find that
\begin{equation}
\partial_t \lambda_\pm = \pm \partial_t \left(
\alpha f^{1/2} / g^{1/2} \right) \: ,
\end{equation}
and using now the evolution equations for $\alpha$ and $g$ we obtain
\begin{equation}
\partial_t \lambda_\pm = \alpha \frac{\lambda_\pm}{g^{1/2}}
\left( 1 - f - \frac{\alpha f'}{2} \right) \tilde{K} \: . 
\end{equation}
In a similar way we find for the spatial derivative
\begin{equation}
\partial_x \lambda_\pm = \lambda_\pm \left[ \frac{D_\alpha}{f} \left( f
+ \frac{\alpha f'}{2} \right) - \frac{D_g}{2} \right] \: . 
\end{equation}
The last two equations together imply that
\begin{eqnarray}
\partial_t \lambda_\pm + \lambda_\pm \partial_x \lambda_\pm &=&
\frac{\alpha}{g^{1/2}} \left[ \left( 1 - f - \frac{\alpha f'}{2} \right)
\left( \tilde{K} \mp \frac{D_\alpha}{f^{1/2}}
\right) \right. \nonumber \\
&\pm& \left. f^{1/2} \left( \frac{D_\alpha}{f} - \frac{D_g}{2} \right)
\right] \: ,
\end{eqnarray}
or in terms of the eigenfields
\begin{eqnarray}
\partial_t \lambda_\pm + \lambda_\pm \partial_x \lambda_\pm = \hspace{35mm}
\nonumber \\
\frac{\alpha}{g^{1/2}} \left[
\left( 1 - f - \frac{\alpha f'}{2} \right)
\omega_\mp  \pm f^{1/2} \omega_0 \right] \: .
\end{eqnarray}

If we now consider a region where \mbox{$\omega_- = \omega_0 = 0$},
then the equation for $\lambda_+$ reduces to
\begin{equation}
\partial_t \lambda_+ + \lambda_+ \partial_x \lambda_+ = 0 \: .
\end{equation}
This is nothing more than Burgers' equation, the prototype for
studying shock formation.  It is easy to understand how this equation
implies genuine shocks: The equation shows that characteristic speeds
are constant along their corresponding characteristic lines, so if
these speeds where not uniform to begin with, and particularly if the
derivative of $\lambda_+$ was initially negative at any point, the
characteristic lines will inevitably cross.

When the characteristics cross the spatial derivative of $\lambda_+$
will become infinite, and as this derivative is given in terms of
eigenfields, the eigenfields will blow up.  This shows that the
blow-ups we studied above correspond to places where the
characteristic lines cross, {\rm i.e.} they are genuine shocks.  The
use of the term ``gauge shocks'' to describe these pathologies is
therefore entirely justified.

\begin{figure}
\epsfxsize=100mm
\epsfysize=70mm
\epsfbox{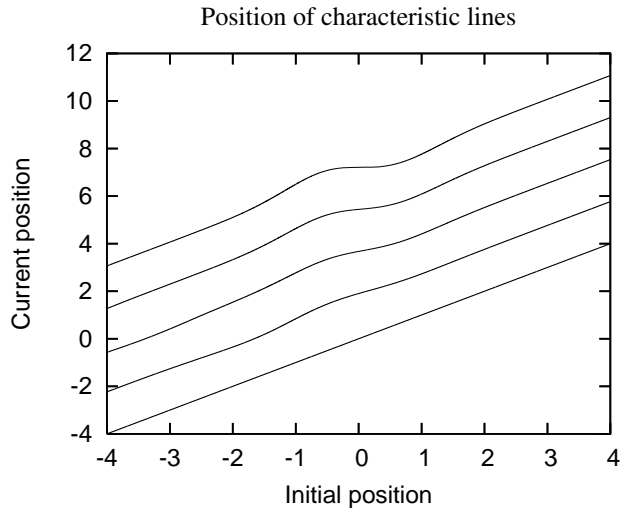}
\caption{Positions of the characteristics during the simulation with
respect to their initial positions, for a series of different times:
$t=0,2.5,5,7.5,10$.  The characteristics were equally spaced a $t=0$
(corresponding to a line at 45 degrees on the plot).  By $t=10$ this
is no longer the case and a plateau has formed indicating that the
characteristics are about to cross.}
\label{fig:charac}
\end{figure}

Figure~\ref{fig:charac} shows the positions of the characteristics
with respect to their initial positions for the simulation discussed
above, at a series of different times: \mbox{$t=0,2.5,5,7.5,10$}.
Notice that for the simulation considered here $f=1/2$ so the
characteristic speed should be close to $\sqrt{f} \sim 0.707$, we
would then expect the lines on the plot to move up by approximately
$0.707 \times 10 = 7.07$.  From the figure we see that even though the
characteristics were equally spaced a $t=0$ (corresponding to a line
at 45 degrees on the plot), at $t=10$ this is no longer the case and a
plateau has formed.  This plateau indicates that those characteristics
initially in the region around the origin are now all essentially at
the same position $x \sim 7$, or in other words, they are about to
cross.

\begin{figure}
\epsfxsize=100mm
\epsfysize=70mm
\epsfbox{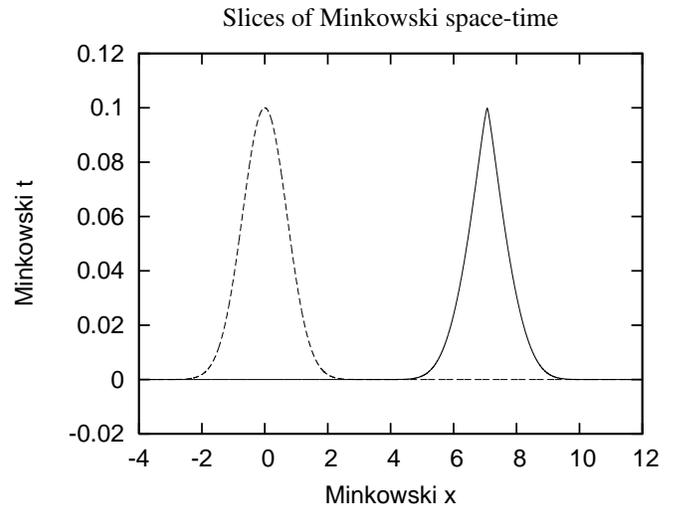}
\caption{Initial hypersurface (dotted line) and final hypersurface at
$t=10$ (solid line), as seen in Minkowski coordinates, for the
simulation discussed in the text.  The final slice has been moved back
in time so that it lies on top of the initial slice at the
boundaries. The initial slice is smooth, while the final slice has a
sharp kink.}
\label{fig:slice}
\end{figure}

One could wonder what the gauge shocks imply about the geometry of the
space-time being evolved.  Since in this case we are simply evolving a
foliation in Minkowski space-time it is clear that the background
geometry remains perfectly regular, the only thing that can become
pathological are the spatial hypersurfaces that determine the
foliation.  Figure~\ref{fig:slice} shows a comparison of the initial
hypersurface and the final hypersurface at $t=10$ as seen in Minkowski
spacetime, using the data from the same numerical simulation discussed
above (for an easier comparison the final slice has been moved back in
time so that it lies on top of the initial slice at the boundaries).
The hypersurfaces are reconstructed by explicitly keeping track of the
position of the normal observers in the original Minkowski coordinates
during the evolution.  Notice how the initial slice is very smooth (it
has in fact a Gaussian profile, see
Appendix~\ref{sec:appendix_initial}), while the final slice has
developed a sharp kink.  This shows that the formation of a gauge
shock indicates that the hypersurface, though still spacelike
everywhere, is no longer smooth (its derivative is now discontinuous).


\section{Conclusions}
\label{sec:conclusions}

I have studied the formation of gauge pathologies associated with the
Bona-Masso family of slicing conditions in simple 1+1 relativity.  The
analysis shown here not only recovers previously known results about
the existence of such pathologies and the conditions necessary for
avoiding them, but also proves that such pathologies arise from the
crossing of the characteristic lines associated with the propagation
of the gauge.  This implies that such pathologies are genuine shocks
and hence the name ``gauge shocks'' used to describe them is entirely
justified.  The gauge shocks appear as discontinuities in the lapse
and spatial metric arising from smooth initial data, and represent
places where the spatial hypersurfaces that determine the foliation of
space-time become non-smooth.

It is also important to point out that Ref.~\cite{Reimann:2004wp}
introduces two different shock avoiding criteria called ``indirect
linear degeneracy'' and the ``source criteria''.  Neither criteria is
fully understood at this point, and they are offered
in~\cite{Reimann:2004wp} more as indicators that seem to work well in
practice than as rigorous conditions necessary for avoiding shocks. In
this manuscript I decided against using either of those criteria
precisely because they are not rigorous.  The purpose here has been to
show that for a simple case (1+1) one does not need to introduce any
{\em ad hoc} criteria, and one can instead show that gauge shocks
develop by directly analyzing the evolution equations and showing
that: 1) blow-ups in the eigenfields do happen (and can even be
predicted in advance), and 2) the characteristic lines associated with
the gauge propagation do cross.


\acknowledgments

I wish to thank to thank Dario Nu\~{n}ez and Bernd Reimann for
many useful discussions and comments. This work was supported in part
by DGAPA-UNAM through grants IN112401 and IN122002.


\appendix

\section{}
\label{sec:appendix_initial}

In this Appendix I will discuss how to construct initial data that
contains only waves propagating in one direction.  Notice that the
simplest way to construct initial data would be to take a trivial
initial slice in Minkowski coordinates, and use a non-trivial initial
lapse function to introduce a later distortion.  However, since in
this case the initial extrinsic curvature vanishes, one sees
from~(\ref{eq:eigenfunc}) that inevitably modes traveling in both
direction will be excited.  We then conclude that in order to have
waves propagating in only one direction one has to consider a
non-trivial initial slice.

The initial data considered here has already been discussed
in~\cite{Alcubierre97a}, and here I just present it for completeness.
We start by considering an initial slice given in Minkowski
coordinates $(t_M,x_M)$ as
\begin{equation}
t_M = h(x_M) \: ,
\end{equation}
with $h$ a profile function that decays rapidly, like a Gaussian
function.  It is then not difficult to show that if we use $x=x_M$ as
our spatial coordinate, the spatial metric an extrinsic curvature turn
out to be
\begin{eqnarray}
g &=& 1 - h'^2  \quad \Rightarrow \quad
D_g = -2 h' h'' / g \: , \\
\qquad K_{xx} &=& - h'' / \sqrt{g} \quad \Rightarrow \quad
\tilde{K} = - h'' / g \; .
\end{eqnarray}

Assume that we want to have waves traveling to the right (the opposite
situation can be done in an analogous way). This means that we want
$\omega_-=0$, which implies
\begin{equation}
D_\alpha = \sqrt{f} \: \tilde{K} = - \sqrt{f} \: h'' / g \; .
\end{equation}
This gives us the following differential equation $\alpha$:
\begin{equation}
\partial_x \alpha = - \alpha \sqrt{f} \: h'' / \left(
1 - h'^2 \right) \; .
\end{equation}
In the particular case when $f$ is a constant the above equation can
be easily integrated to find
\begin{equation}
\alpha = \left( \frac{1 - h'}{1 + h'} \right)^{\sqrt{f}/2} \; .
\end{equation}
From this one can reconstruct the rescaled eigenfunction $\Omega_+$
and use Eq.~(\ref{eq:blowuptime}) to predict the time of shock
formation given a specific form of $h(x)$.

In the simulation shown in Sec.~\ref{sec:shocks} above, the profile
function $h$ was taken to be a simple Gaussian of the form
\begin{equation}
h(x) = e^{- x^2} / 10 \: .
\end{equation}


\section{}
\label{sec:appendix_bona}

Here I will briefly counter the arguments given by Bona {\em et al.}
in~\cite{Bona:2004ma} to claim that gauge shocks are, in their words,
completely discarded.  Their main argument is based on looking at the
so-called ``foliation equation'', that looks at the foliation as the
evolution of a scalar function $T$ whose level sets correspond to the
3+1 slices.  In~\cite{Alcubierre02b} I showed that the Bona-Masso
slicing condition can be written in covariant form as
\begin{equation}
\left[ g^{\mu \nu} + \left( 1 - \frac{1}{f(\alpha)} \right)
n^\mu n^\nu \right] \nabla_\mu \nabla_\nu T  = 0 \; ,
\label{eq:foliation}
\end{equation}
where $n^\mu$ is the unit normal vector to the hypersurfaces
\begin{equation}
n_\mu = -\alpha \nabla_\mu T  \; ,
\end{equation}
and with the lapse function $\alpha$ given in terms of $T$ as
\begin{equation}
\alpha = \left( - \nabla_\mu T \; \nabla^\mu T \right)^{-1/2} \; .
\end{equation}
Assume for a moment that $f=1$. In that case Eq.~(\ref{eq:foliation})
reduces to the standard wave equation.  Consider now a specific point
in spacetime and use locally flat coordinates.  It is clear that in
such coordinates the foliation equation has constant characteristic
speeds equal to 1.  This implies that the characteristic lines do not
cross, and if the lines do not cross in some set of regular
coordinates they won't cross in any regular coordinates.  We then
conclude that in this case there are no shocks.

But what happens when $f \neq 1$?  Bona {\em et al.} argue that in
that case one can always start with an arbitrary regular slice such
that $T={\rm constant}$ on that slice.  From that one can easily see
that, as long as the slice is spacelike, one can always recover all
second partial derivatives of $T$ from equation~(\ref{eq:foliation}),
which allows one to construct the next hypersurface.  This is
certainly true, but it does not follow from here that one can continue
this procedure for any finite time, since all we have done is show
that given regular initial data there is locally a regular solution of
the foliation equation, but shocks are precisely solutions that fail
after a {\em finite time}.

In~\cite{Alcubierre02b} I showed that if one takes
Eq.~(\ref{eq:foliation}) with $f \neq 1$ and considers locally flat
coordinates, the characteristic speeds are {\em not} constant anymore.
There is then no guarantee that they will not cross after a finite
time.  In fact, in~\cite{Alcubierre02b} it is shown that if one
applies the standard analysis coming from the theory of PDE's to
Eq.~(\ref{eq:foliation}) one finds that shocks will form unless
condition~(\ref{eq:shockcondition}) holds.  The problem with the
argument of Bona {\em et al.} is therefore that they failed to
consider that when one deals with shocks, smooth initial data always
guarantees smooth solutions {\em locally}, but not after a {\em finite
time}.

Bona {\em et al.} also present other arguments. For example,
starting from their equation (6) for the speed of light
\begin{equation}
c = \frac{dl}{dt} = \pm \sqrt{ \gamma_{ij} \frac{dx^i}{dt}
\frac{dx^j}{dt} } = \pm \alpha \; ,
\end{equation}
they obtain the shock avoiding condition $f=k/\alpha^2$ with $k$
constant, which is clearly different from~(\ref{eq:alphanoshock}).
But in fact the last equality in the equation above is wrong, the
speed of light is not equal to the lapse in a general coordinate
system, one is missing a factor of $\sqrt{\gamma^{ii}}$ (with $x^i$
the direction of propagation).  Inserting this factor takes us back to
the shock avoiding condition~(\ref{eq:alphanoshock}).

Bona {\em et al.} also consider cosmological-type scenarios
(homogeneous and isotropic), and derive different conditions for
avoiding blow-ups in the lapse (see equation (18)
of~\cite{Bona:2004ma}).  But since in this case there is no
propagation and just growth in place, the whole concept of shocks and
characteristic speeds is simply non-applicable.


\bibliographystyle{bibtex/apsrev} \bibliography{bibtex/referencias}


\end{document}